\documentclass[proceedings]{JHEP3} 
\PrHEP{}
\conference{International Workshop on Astroparticle and High Energy
Physics}

\usepackage{epsfig,multicol}	
\title{Recent Results and Status of TEXONO Experiments}
\author{\speaker{Venktesh~Singh} and Henry T.~Wong\\
{\large (on behalf of the TEXONO Collaboration)}\\
Institute of Physics, Academia Sinica, Taipei 11529, Taiwan\\
E-mail:\email{vsingh@phys.sinica.edu.tw},\email{htwong@phys.sinica.edu.tw}
}
\conference{AHEP03}
	\abstract{
This article reviews
the research program and efforts for the
TEXONO Collaboration on neutrino
and astro-particle physics.
The ``flagship'' program is on reactor-based neutrino physics
at the Kuo-Sheng (KS) Power Plant in Taiwan.
A limit on the neutrino magnetic moment of
$\rm{ \munuebar < 1.3 \times 10^{-10} ~ \mub}$
at 90\%  confidence level was derived from
measurements with a high purity germanium detector.
Other physics topics at KS, as well as 
the various R\&D program, are discussed.
}

\def\nuebar{\bar{\nu_e}}
\def\nue{\nu_e}
\def\munu{\mu_{\nu}}
\def\gammanu{\Gamma_{\nu}}
\def\dm2{\rm{\Delta m^2}}

\def\s2tw{\rm{ sin ^2 \theta _W }}
\def\am241{\rm{ ^{241} Am }}
\def\u238{\rm{ ^{238} U }}
\def\th232{\rm{ ^{232} Th }}
\def\k40{\rm{ ^{40} K }}
\def\th232{\rm{ ^{232} Th }}
\def\u238{\rm{ ^{238} U }}
\def\cs137{\rm{^{137} Cs }}
\def\ba133{\rm{^{133} Ba }}

\def\s2tw{\rm{ sin ^2 \theta _W }}
\def\munuebar{\rm{\mu_{\nuebar}}}
\def\mub{\rm{\mu_B}}

\def\ke10{\rm{\kappa_e}}

\begin{document}

\section{Introduction and History}

The 
TEXONO\footnote{{\bf T}aiwan {\bf EX}periment {\bf O}n {\bf N}eutrin{\bf O}}
Collaboration\cite{texono} has been built up since 1997 to
initiate and pursue an experimental
program in Neutrino and Astroparticle Physics\cite{start}.
The Collaboration comprises
more than 40 research scientists from
major institutes/universities
in Taiwan (Academia Sinica$^{\dagger}$, Chung-Kuo
Institute of Technology, Institute of Nuclear
Energy Research, National Taiwan University, National Tsing Hua
University, and Kuo-Sheng Nuclear Power Station),
China (Institute of High Energy Physics$^{\dagger}$,
Institute of Atomic Energy$^{\dagger}$,
Institute of Radiation Protection,
Nanjing University, Tsing Hua University)
and the United States (University of Maryland),
with AS, IHEP and IAE (with $^{\dagger}$)
being the leading groups.
It is the first research collaboration 
of this size and magnitude among 
scientists from Taiwan and China\cite{sciencemag}.

Results from recent neutrino experiments
strongly favor neutrino oscillations 
which imply neutrino masses and 
mixings~\cite{pdg}.
Their physical origin and experimental consequences
are not fully understood.
There are strong motivations for
further experimental efforts to 
shed light on these fundamental questions 
by probing standard and anomalous neutrino properties
and interactions.  The results
can constrain theoretical models
which will be necessary to interpret
the future precision data.
In addition, these studies 
would also explore new detection channels to
open up new avenues of investigations. 

The TEXONO research program is based on the
the unexplored and unexploited theme of adopting
detectors with
high-Z nuclei, such as solid state device and
scintillating crystals,
for low-energy low-background experiments
in Neutrino and Astroparticle Physics\cite{prospects}.
The main effort is 
a reactor neutrino experiment
at the Kuo-Sheng (KS) Nuclear Power Station
in Taiwan to study low energy neutrino
properties and interactions\cite{program}.
The Kuo-Sheng experiment is the first particle 
physics experiment in Taiwan.
In parallel to the reactor experiment,
various R\&D efforts coherent with the
theme are initiated and pursued. 

Subsequent sections highlight the results 
and  status of the program.

\section{Kuo-Sheng Neutrino Laboratory}

\begin{figure}
\center
{\bf a)}
\epsfig{figure=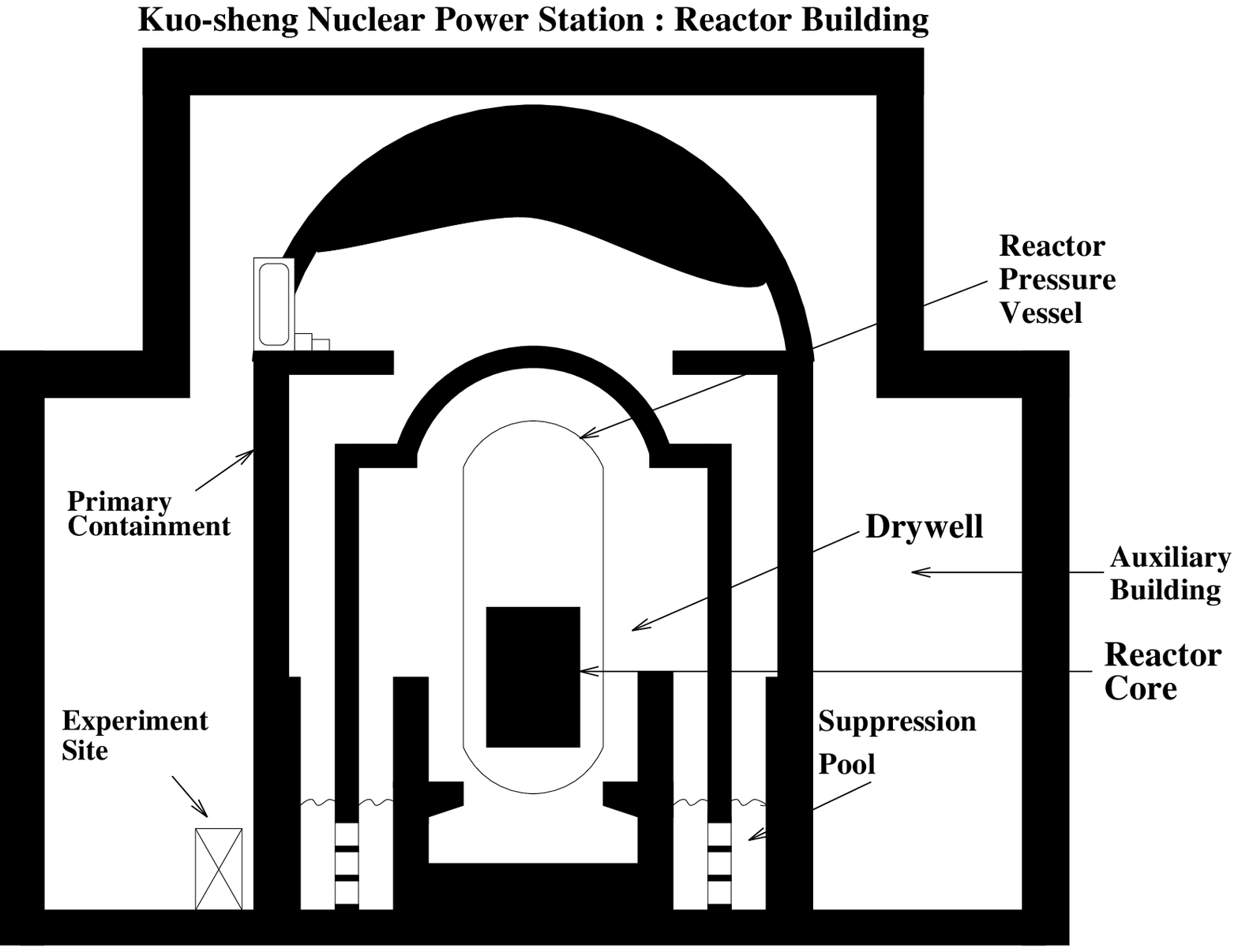,height=7cm,angle=270}
{\bf b)}
\epsfig{figure=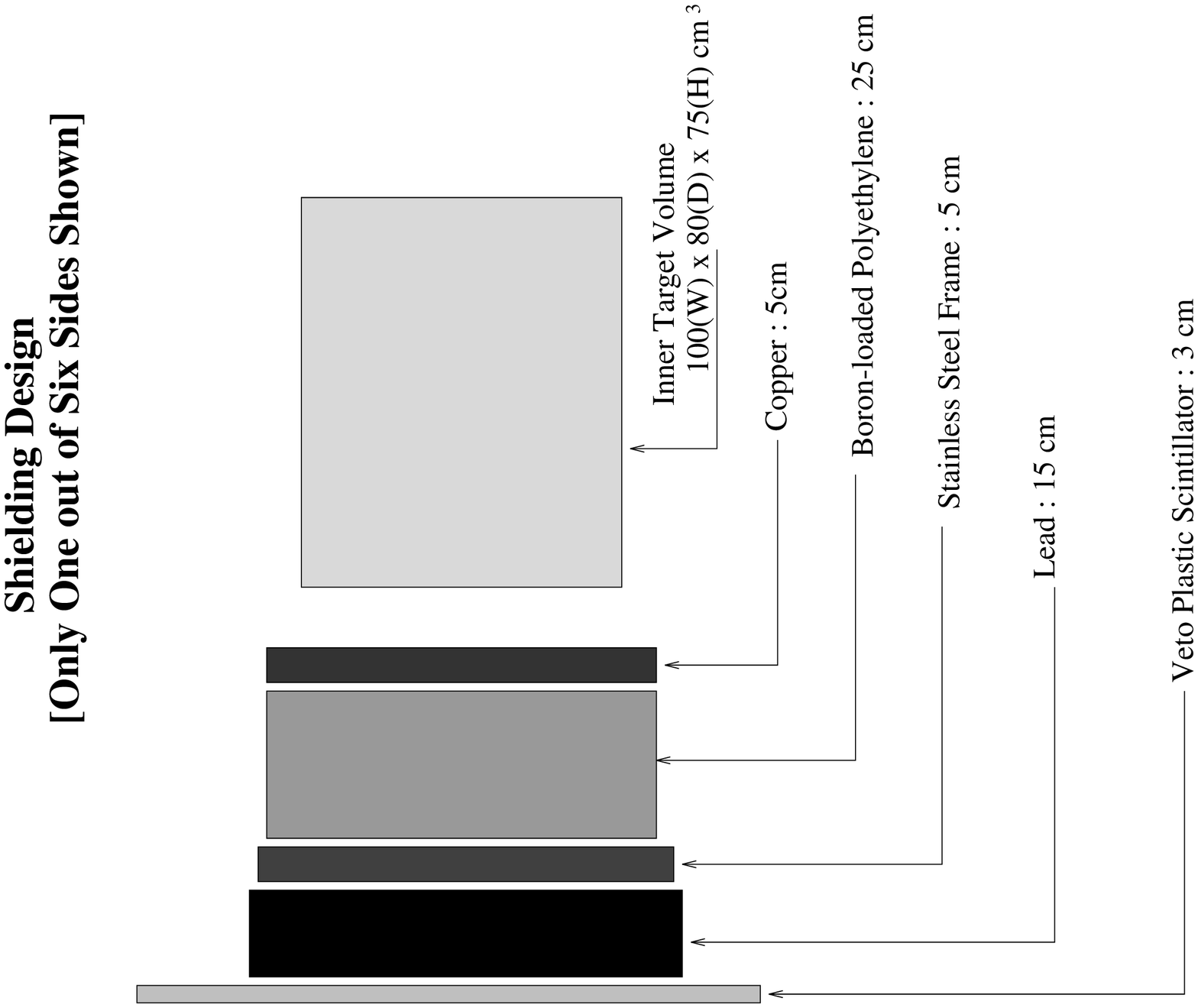,height=5cm,angle=270}
\caption{
(a) Schematic side view, not drawn to scale,
of the Kuo-Sheng Nuclear Power Station
Reactor Building,
indicating the experimental site.
The reactor core-detector distance is about 28~m.
(b) Schematic layout of the inner target space,
passive shieldings and cosmic-ray veto panels.
The coverage is 4$\pi$ but only one face is shown.
}
\label{ksnpssite}
\end{figure}

The ``Kuo-Sheng Neutrino Laboratory''
is located at a distance of 28~m from the core \#1
of the Kuo-Sheng Nuclear Power Station 
at the northern shore of Taiwan\cite{program}.
A schematic view is depicted in Figure~\ref{ksnpssite}a.

A multi-purpose ``inner target'' detector space of
100~cm$\times$80~cm$\times$75~cm is
enclosed by 4$\pi$ passive shielding materials
which have a total weight of 50 tons.
The shielding provides attenuation
to the ambient neutron and gamma background, and
consists of, from inside out,
5~cm of OFHC copper, 25~cm of boron-loaded
polyethylene, 5~cm of steel, 15~cm of lead,
and cosmic-ray veto scintillator panels.
The schematic layout of one side
is shown in Figure~\ref{ksnpssite}b.

Different detectors can be placed in the
inner space for the different scientific goals.
The detectors are read out by a versatile
electronics and data acquisition systems\cite{eledaq}
based on a 16-channel, 20~MHz, 8-bit 
Flash Analog-to-Digital-Convertor~(FADC)  module.
The readout allows full recording of all the relevant pulse
shape and timing information for as long as several ms
after the initial trigger.
Software procedures have been devised to extend the 
effective dynamic range from the  nominal 8-bit
measurement range provided by the FADC\cite{dyrange}.
The reactor laboratory is connected via telephone line to
the home-base laboratory at AS, where remote access 
and monitoring are performed regularly. Data are stored
and accessed via the PC IDE-bus 
from a cluster of multi-disks arrays 
each with 800~Gbyte of memory.

\begin{figure}
\center
\epsfig{figure=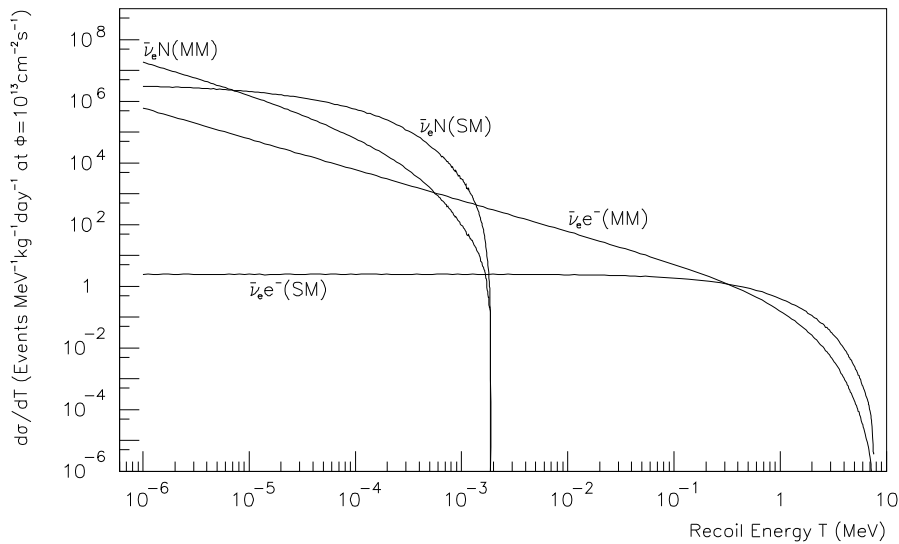,height=8cm}
\caption{
Differential cross section showing the
recoil energy spectrum in
$\nuebar$-e and coherent $\nuebar$-N
scatterings,
at a reactor neutrino flux of
$\rm{10^{13}~cm^{-2} s^{-1}}$,
for the Standard Model (SM) processes and
due to a neutrino
magnetic moment (MM) of 10$^{-10}~\mub$.
}
\label{spectra}
\end{figure}

The measure-able nuclear and electron recoil spectra
due to reactor $\nuebar$ are depicted in Figure~\ref{spectra},
showing the effects due to 
Standard Model [$\rm{\sigma (SM) }$] and
magnetic moment [$\rm{\sigma ( \munu ) }$] $\nuebar$-electron
scatterings\cite{sigmanue}, as well as the Standard Model neutrino
coherent scatterings on the nuclei [$\rm{\sigma (coh) }$].
It was recognized recently\cite{sensit} that
due to the uncertainties in the modeling of the
low energy part of the reactor neutrino spectra,
experiments to measure $\rm{\sigma (SM) }$
with reactor neutrinos
should focus on higher electron recoil
energies (T$>$1.5~MeV), while
$\munu$ searches should base
on measurements with T$<$100~keV.
Observation of $\rm{\sigma (coh) }$ would
require detectors with sub-keV sensitivities.

Accordingly, data taking were optimized for
with these strategies.
An ultra low-background high purity germanium (ULB-HPGe)
detector was used for Period I (June 2001 till May 2002) 
data taking, while 186~kg of CsI(Tl) crystal scintillators
were added in for Period II (starting January 2003).
Both detector systems operate in parallel 
with the same data acquisition system 
but independent triggers.
The target detectors are housed in a nitrogen
environment to prevent background events due to the
diffusion of the radioactive radon gas.

\subsection{Germanium Detector}

\begin{figure}
\center
{\bf a)}
\epsfig{figure=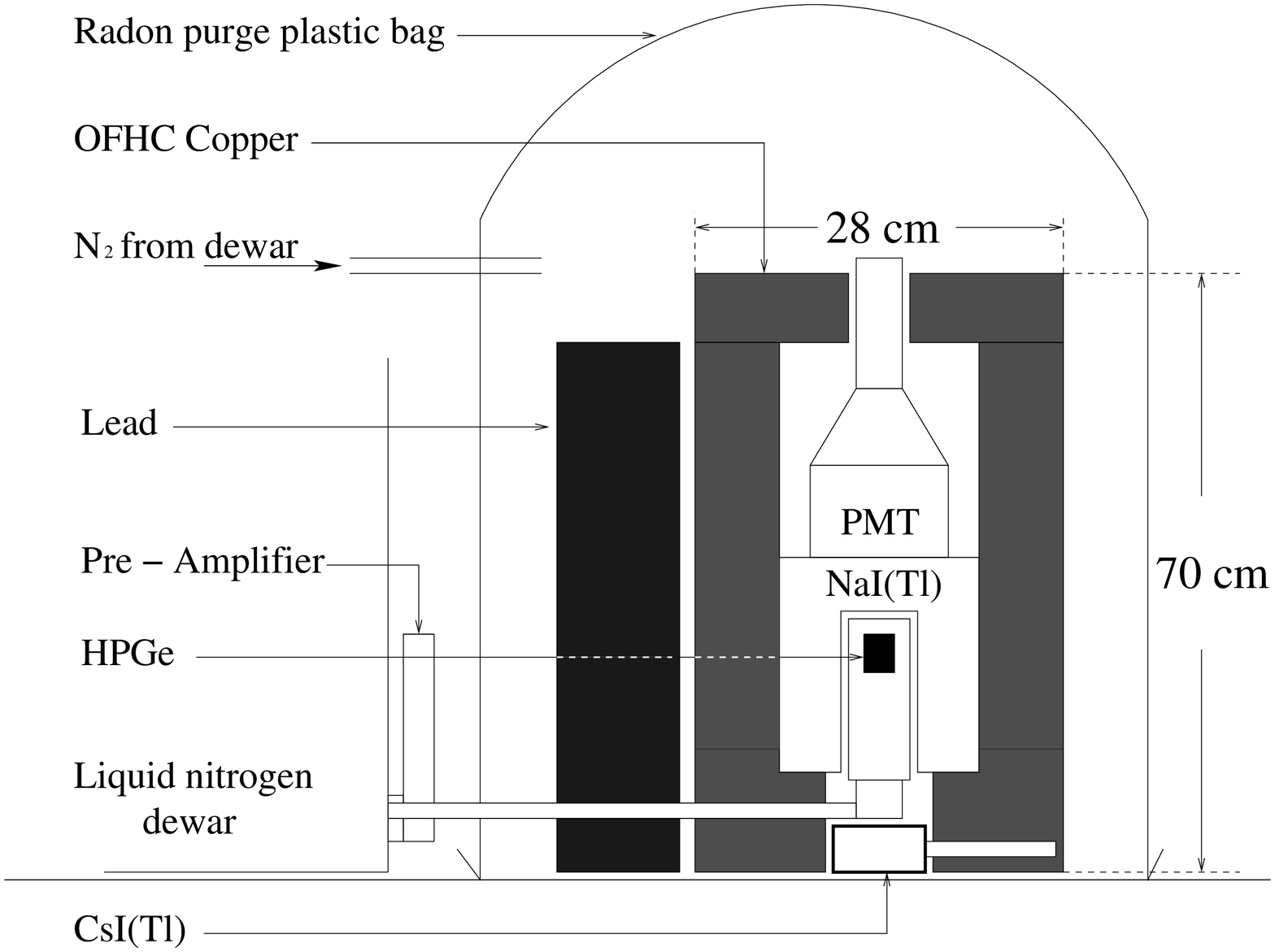,height=5.5cm}
{\bf b)}
\epsfig{figure=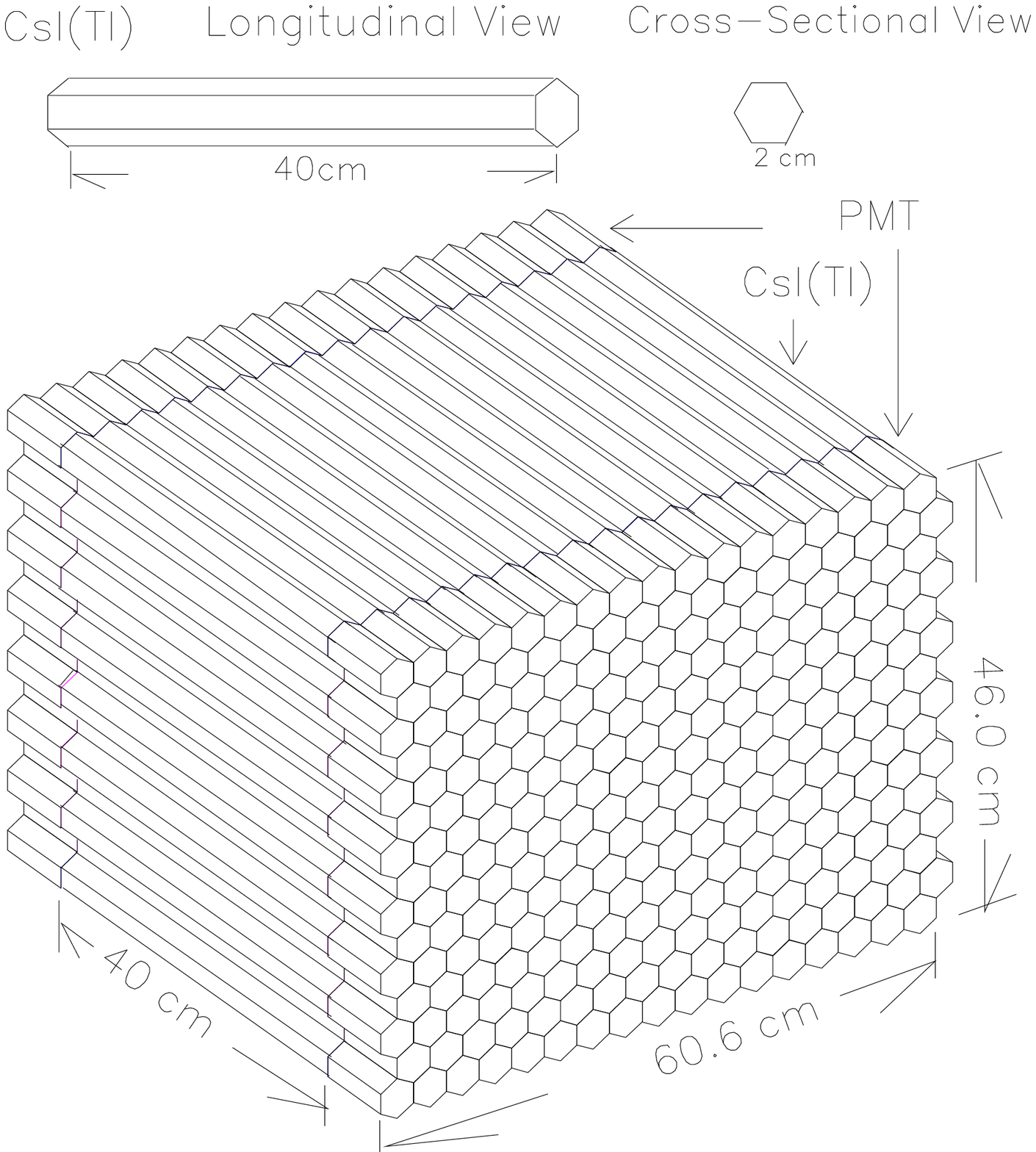,height=5.5cm}
\caption{
Schematic drawings of 
(a) the ULB-HPGe
detector with its anti-Compton scintillators
and passive shieldings; and
(b) the CsI(Tl) target configuration where
a total of 93 modules (186~kg)
is installed for Period II.
}
\label{target}
\end{figure}

As depicted in Figure~\ref{target}a,
the ULB-HPGe is surrounded by NaI(Tl) and CsI(Tl) crystal
scintillators as anti-Compton detectors, and the whole set-up
is further enclosed by another 3.5~cm of OFHC copper and
lead blocks.

\begin{figure}
\center
\epsfig{figure=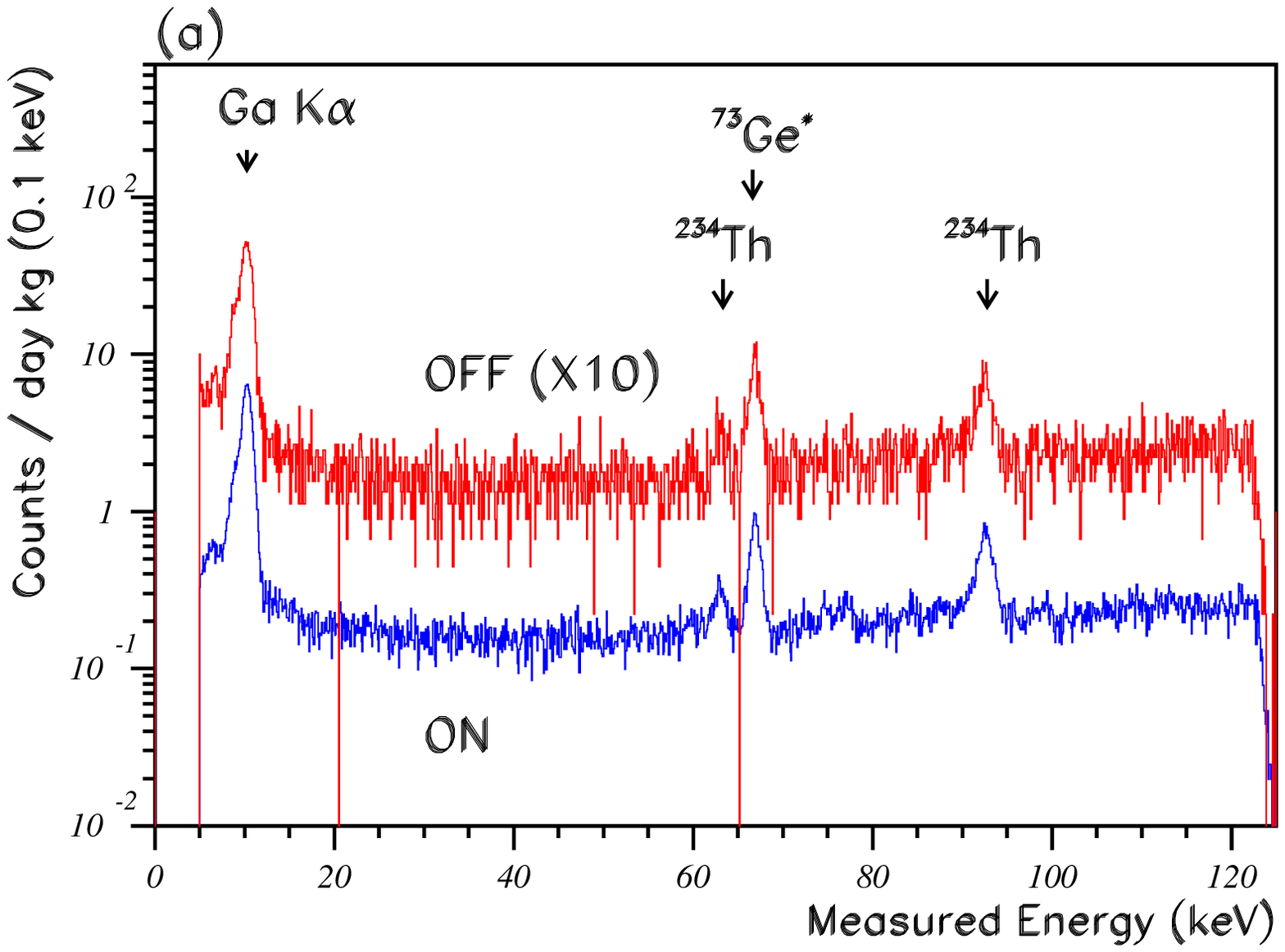,height=5.0cm}
\epsfig{figure=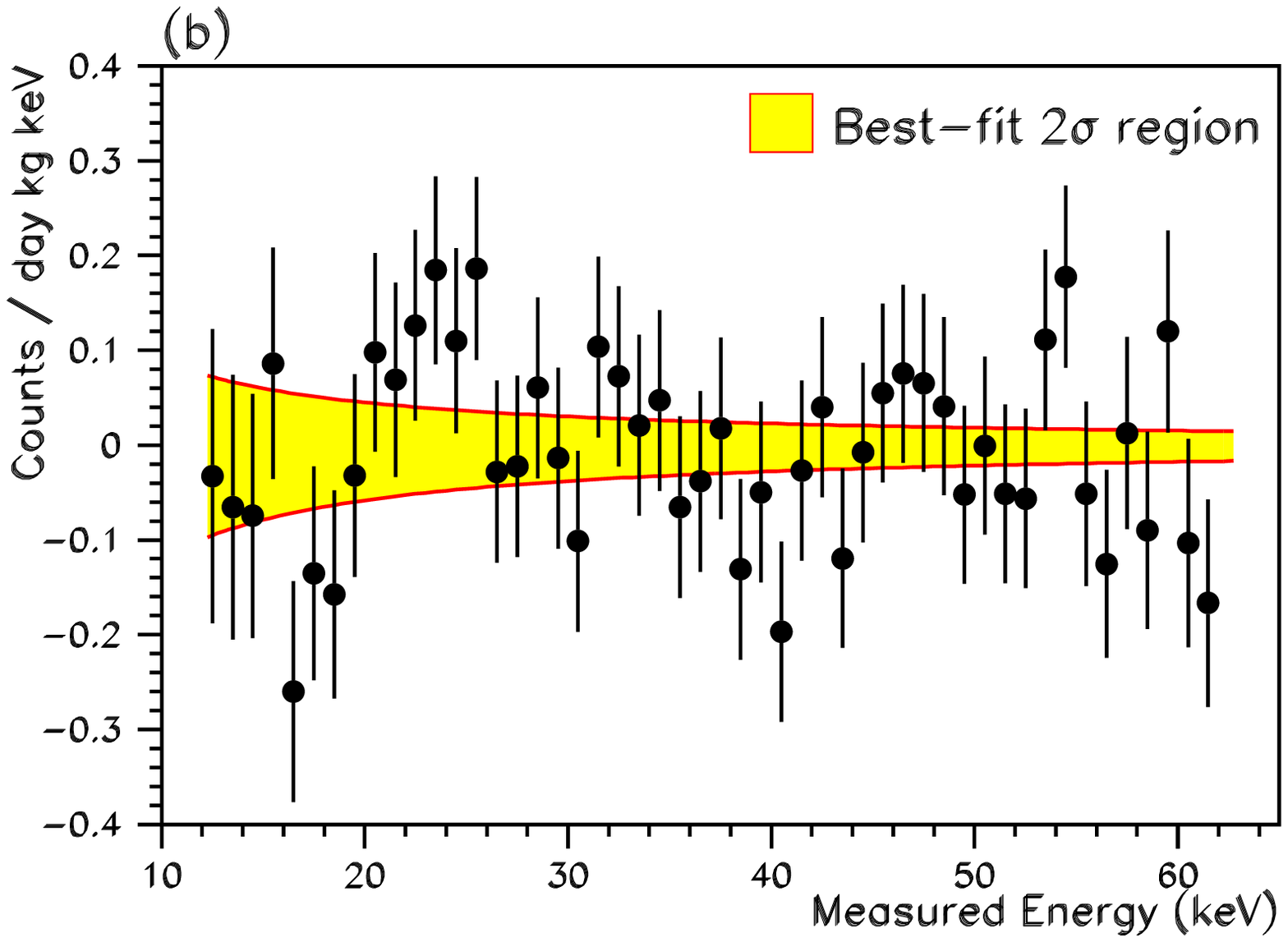,height=5.0cm}
\caption{
(a)
The ON and OFF spectra after all identifiable
background suppressed, for 4720 and 1250 hours
of data, respectively.
(b)
The residual of the ON spectrum over
the OFF background, with the 
2-$\sigma$ best-fit region overlaid.
}
\label{geresults}
\end{figure}

After suppression of cosmic-induced 
background, anti-Compton vetos and convoluted events
by pulse shape discrimination,
the measured spectra for 4712/1250 hours of 
Reactor ON/OFF data in Period I\cite{munupaper} are displayed in
Figure~\ref{geresults}a.
Background at the range of 1~keV$^{-1}$kg$^{-1}$day$^{-1}$
and a detector threshold of 5~keV are  achieved.
These are the levels comparable to underground Dark Matter
experiment. 
Comparison of the ON and OFF spectra shows no excess
and limits of the neutrino magnetic moment 
$\rm{ \munuebar < 1.3(1.0) \times 10^{-10} ~ \mub}$
at 90(68)\% confidence level (CL) were set.
The residual plot together with the best-fit
regions are depicted in Figure~\ref{geresults}b.

\begin{figure}
\center
\epsfig{figure=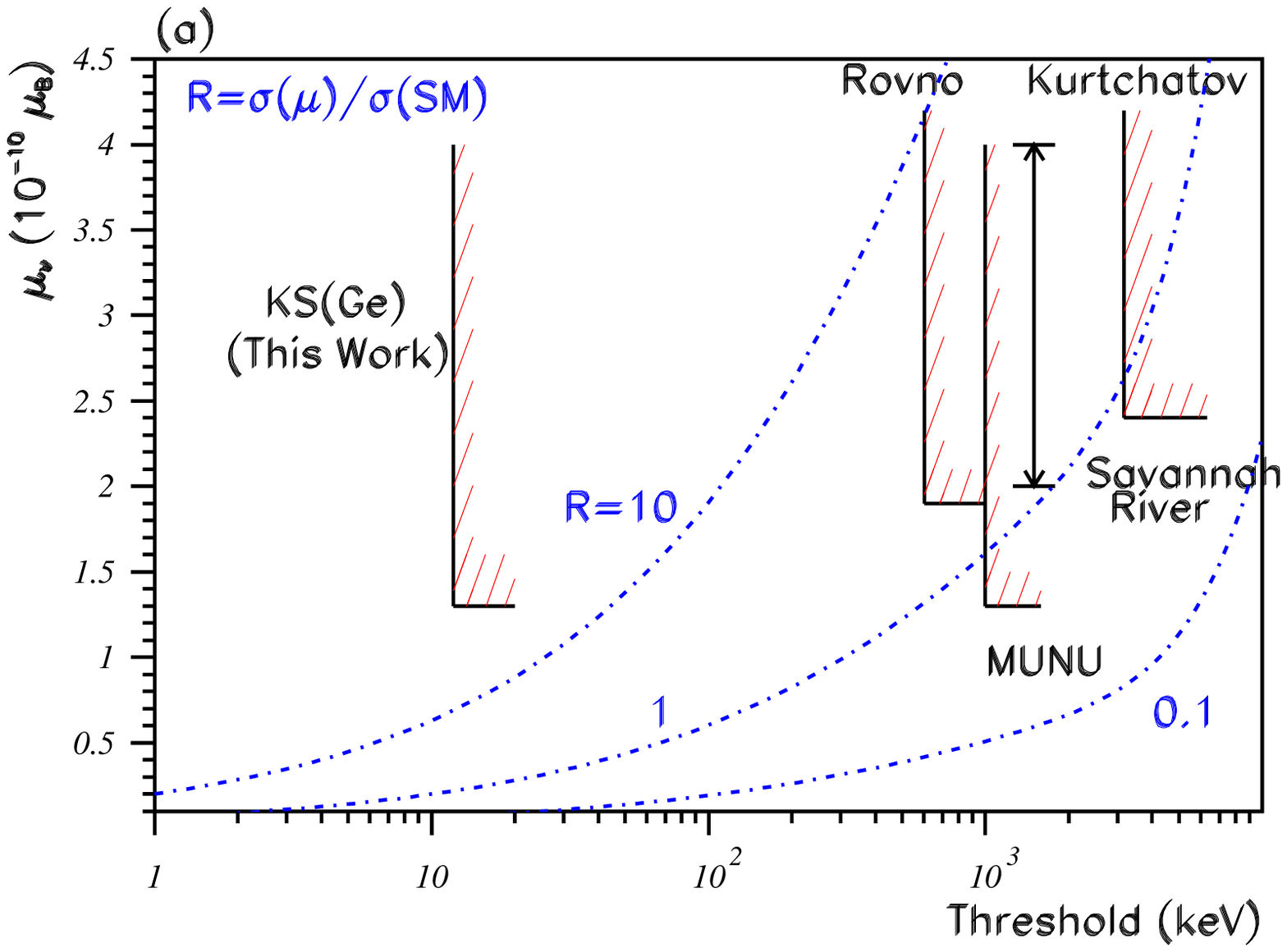,height=5.0cm}
\epsfig{figure=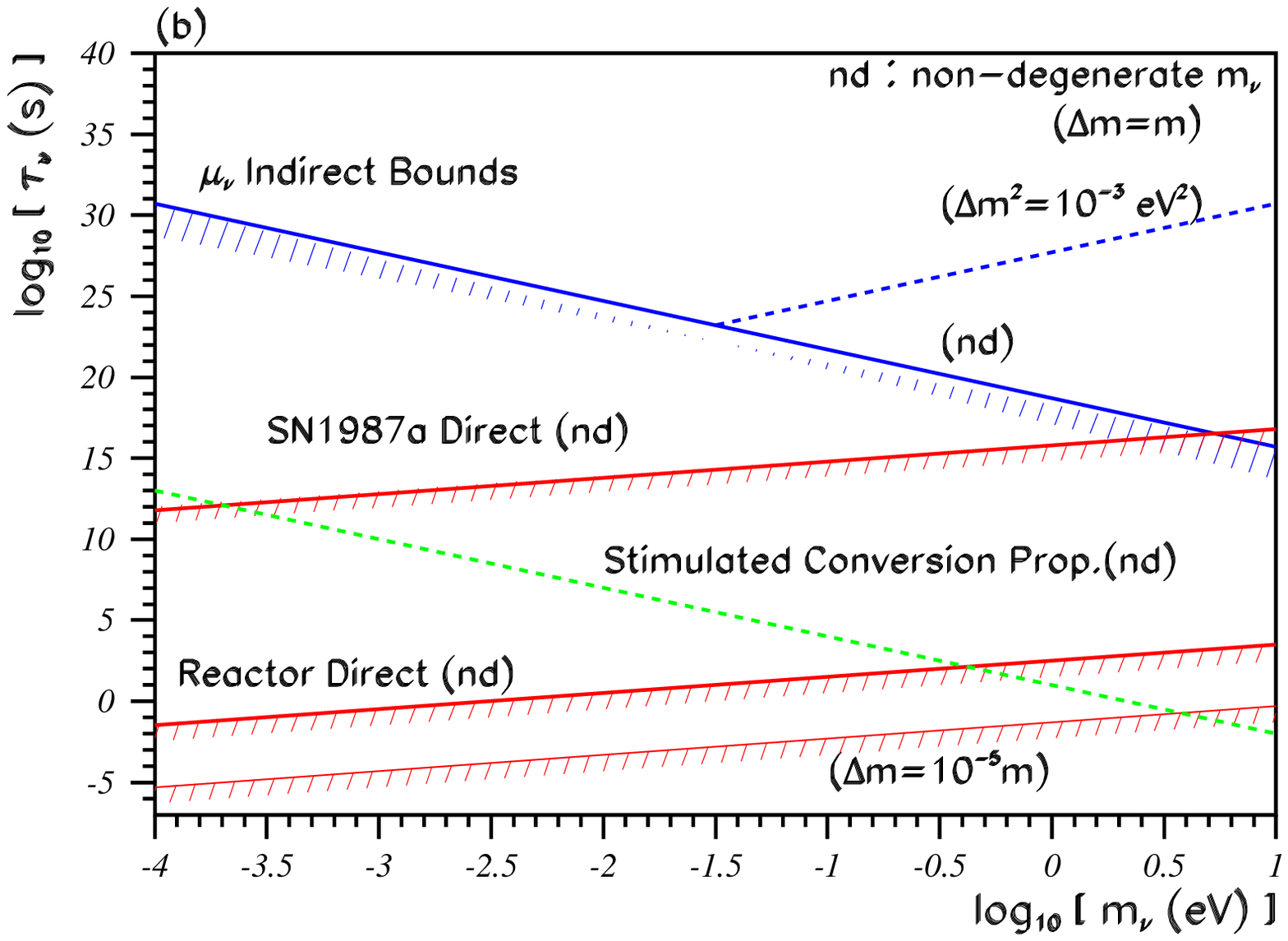,height=5.0cm}
\caption{
Summary of the results in
(a) the searches of neutrino
magnetic moments with reactor neutrinos,
and
(b) the bounds of
neutrino radiative decay lifetime.
}
\label{summaryplots}
\end{figure}

Depicted in Figure~\ref{summaryplots}a is the
summary of the results in $\rm{\munuebar}$ searches
versus the achieved threshold
in reactor experiments.
The dotted lines denote the
$\rm{R = \sigma ({\mu}) / \sigma (SM) }$ ratio at a
particular (T,$\rm{\munuebar}$).
The KS(Ge) experiment has
a much lower threshold of 12~keV compared
to the other measurements.
The large R-values
imply that the KS results
are robust against the uncertainties in the
SM cross-sections. 
The neutrino-photon couplings
probed by $\munu$-searches in
$\nu$-e scatterings are related to 
the neutrino radiative decays ($\gammanu$)\cite{rdk}.
The indirect bounds on $\gammanu$ can be inferred 
and displayed in Figure~\ref{summaryplots}b. 
It can be seen that $\nu$-e scatterings give much more
stringent bounds than the direct approaches.

The KS data with ULB-HPGe are the lowest threshold data 
so far for reactor neutrino experiments, and therefore 
allow the studies of several new and more speculative topics. 
Nuclear fission at reactor cores also produce 
electron neutrino ($\nue$) through the 
production of unstable isotopes, such as
$^{51}$Cr and $^{55}$Fe, via neutron capture,
The subsequent decays of these isotopes by electron capture
would produce mono-energetic $\nue$. 
A realistic neutron transfer simulation
has been carried out to estimate the flux. 
Physics analysis on the $\munu$ and $\gammanu$ of $\nue$
will be performed, while the potentials for other 
physics applications will be studied.
In additional, an {\it inclusive} analysis of
the anomalous neutrino interactions with matter 
will be performed.

\subsection{Scintillating CsI(Tl) Crystals}

The potential merits of crystal scintillators
for low-background low-energy experiments were
recently discussed\cite{prospects}.
The CsI(Tl) detector configuration 
for the KS experiment 
is displayed in Figure~\ref{target}b.
Each crystal module
is 2~kg in mass and 
consists of a hexagonal-shaped cross-section with 2~cm
side and a length 40~cm. 
The light output are read out
at both ends  by custom-designed 29~mm diameter
photo-multipliers (PMTs) with low-activity glass. 
The sum and difference of the PMT signals gives information
on the energy and the longitudinal position of
the events, respectively.
A total of 186~kg (or 93~modules) have been
commissioned for the Period II data taking.
A major physics goal is the measurement of 
the Standard Model neutrino-electron
scattering cross sections. 
The strategy\cite{sensit} is
to focus on data at high ($>$2~MeV) recoil
energy. The large target mass compensates the
drop in the cross-sections.

\begin{figure}
\center
\epsfig{file=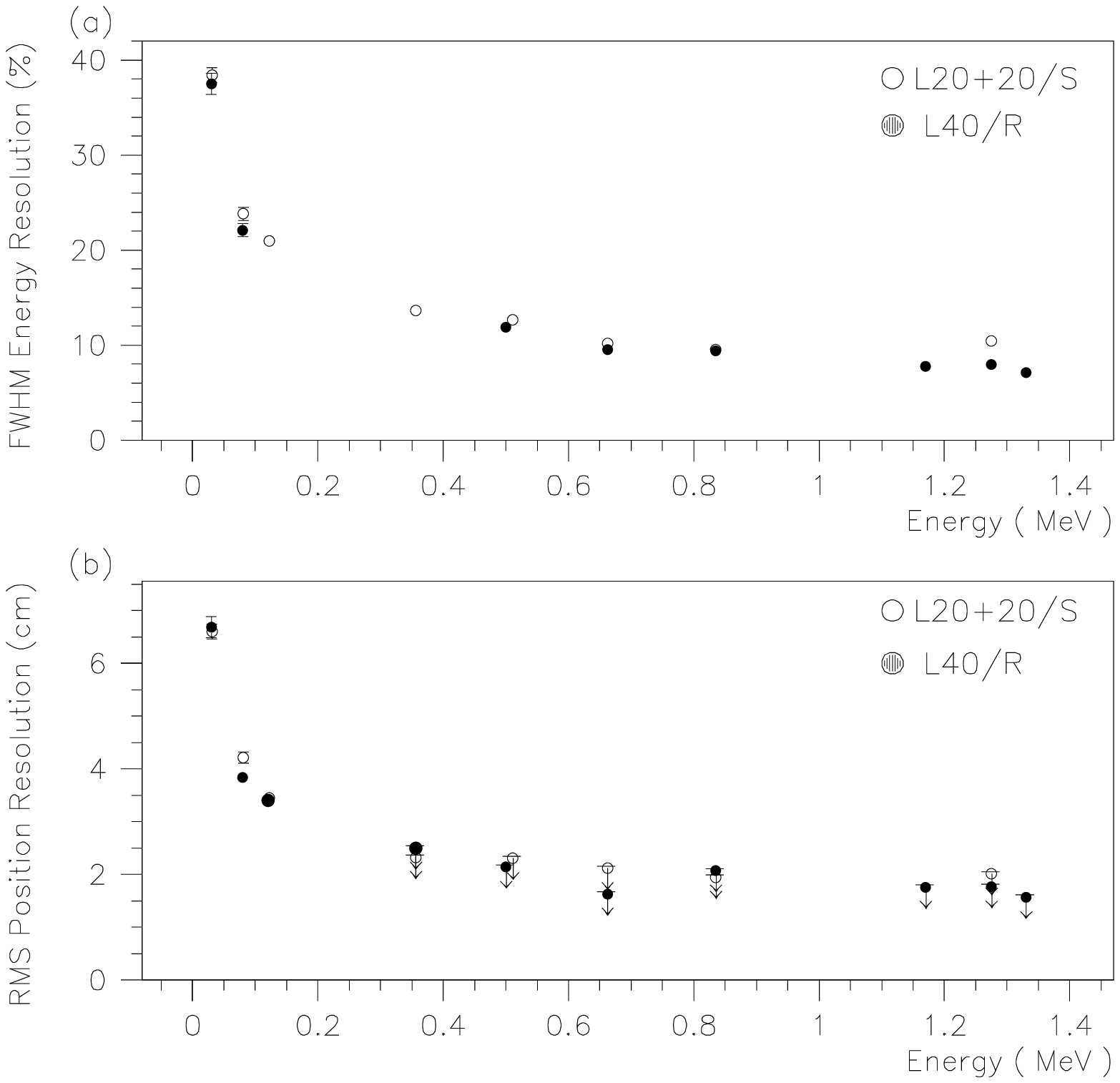,width=8.5cm}
\epsfig{file=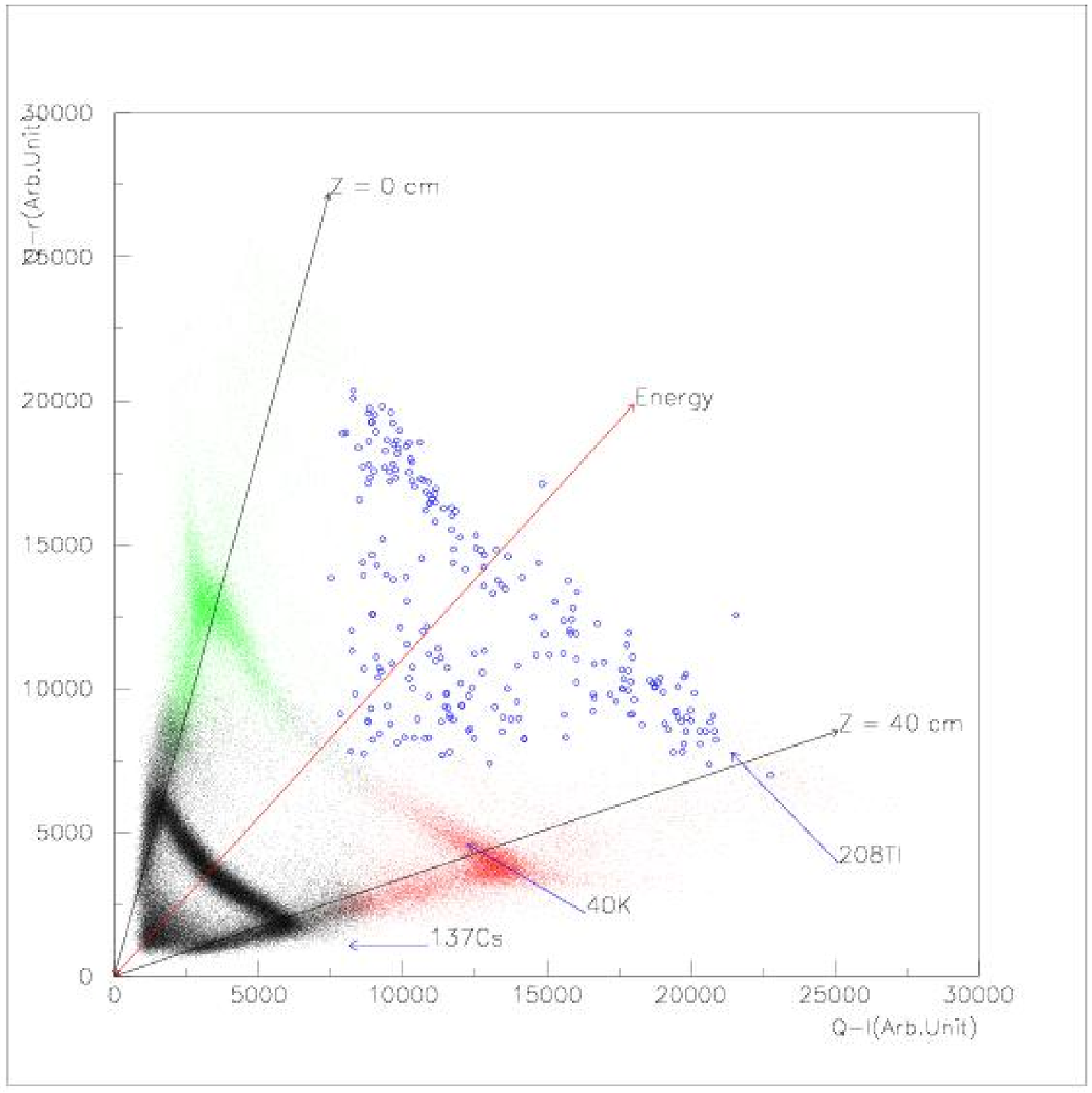,width=6.1cm}
\caption{
The variation of
(a) FWHM energy resolution
and  (b) RMS position resolution
with energy
for the CsI(Tl) crystal modules.
Only upper limits are shown
for the higher energy points in (b)
since the events are not localized.
(c) $\rm{Q_L}$ versus $\rm{Q_R}$ distributions
for single site events.
}
\label{csiresults}
\end{figure}

Extensive measurements on the crystal prototype
modules have been performed\cite{proto}.
The energy and spatial resolutions as
functions of energy are  depicted
in Figure~\ref{csiresults}a 
and \ref{csiresults}b, respectively.
The energy is defined by the total
light collection  $\rm{\sqrt{Q_L \ast Q_R}}$.
It can be seen that
a $\sim$10\% FWHM energy 
resolution is achieved at 660~keV.
The detection threshold (where signals
are measured at both PMTs) is $<$20~keV.
The longitudinal position can be obtained
by considering the variation of the ratio
$\rm { R = (  Q_L - Q_R ) / (  Q_L + Q_R ) }$
along the crystal.
Resolutions of $\sim$2~cm and $\sim$3.5~cm at
660~keV and 200~keV, respectively, have been demonstrated.

In addition, CsI(Tl) provides powerful
pulse shape discrimination (PSD) capabilities
to differentiate $\gamma$/e from $\alpha$ events,
with an excellent separation of $>$99\% above 500~keV.
The light output for $\alpha$'s in CsI(Tl) is quenched
less than that in liquid scintillators.
The absence of multiple $\alpha$-peaks 
above 3~MeV~\cite{csibkg} in the prototype measurements
suggests that a $^{238}$U and $^{232}$Th
concentration (assuming equilibrium)
of $< 10^{-12}$~g/g can be achieved.
It has been shown that PSD can also be
achieved for pulse shapes which are partially
saturated\cite{dyrange}.

The typical $\rm{Q_L}$ versus
$\rm{Q_R}$ distributions for single-site
events after cosmic vetos
are depicted in Figure~\ref{csiresults}c.
There are evidence of contamination of
internal radioactivity due to residual $^{137}$Cs,
such that the distributions of the
662-keV events are uniform across
the 40~cm crystal length.
Events due to $\gamma$-background
from $^{40}$K (1460~keV) and $^{208}$Tl (2612~keV),
on the other hand, occur more frequently  
near both edges, indicating that they are
from sources external to the crystals.
The background is very low above 2.6~MeV,
making this a favorable range to provide
a measurement of $\rm{\sigma ( SM )}$.

\begin{figure}
\center
\epsfig{file=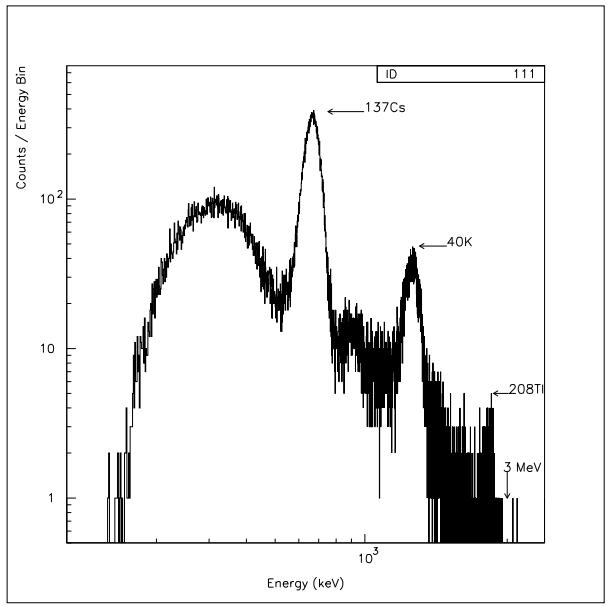,width=10cm}
\caption{Background energy spectrum of recoil electrons.
}
\label{background}
\end{figure}
A preliminary background energy spectrum of recoil electrons
for central crystal and only one week OFF - period data sets are 
depicted in Figure~\ref{background}. For this spectrum, initially 
we applied moderate cuts like single event hit, single crystal 
hit and cosmic ray cuts. After applying these cuts we were able 
to see $^{137}$Cs and $^{40}$K peaks. Further  
we applied a few specific cuts such as Alpha event cut and 
accidental event cut. Alpha events have very fast fall time 
compared to normal event and they are all located 
around 2.0~MeV. After applying these cuts the 
third peak of $^{208}$Tl could be seen. 
In this spectrum we have not used ${z}$-position
cuts. From this spectrum we can see that above 2.9~MeV there are 
only TWO data points and there is no data point above 3.5~MeV.
Thus, the signal to background ratio is 1/17 above 3.0~MeV and 
this ratio must be better when we calculate in term of ON 
period data over OFF period data and applying ${z}$-position cut. 

\section{R\&D Program}

Various R\&D projects with
are proceeding in parallel to the 
KS reactor neutrino experiment. 
The highlights are:

\subsection{Low Energy Neutrino Detection}

It is recognized recently that $^{176}$Yb and $^{160}$Gd are
good candidate targets in the detection of
solar neutrino ($\nue$) by providing a flavor-specific
time-delayed tag\cite{lens}.
Our work
on the Gd-loaded scintillating crystal GSO\cite{gso}
indicated major background issues to be
addressed.
We are exploring the possibilities
of developing Yb-based scintillating crystals, like
doping the known 
crystals $\rm{Yb Al_{l5} O_{12}}$(YbAG) and
$\rm{Yb Al O_{3}}$(YbAP) with scintillators.

\begin{figure}
\center
\epsfig{file=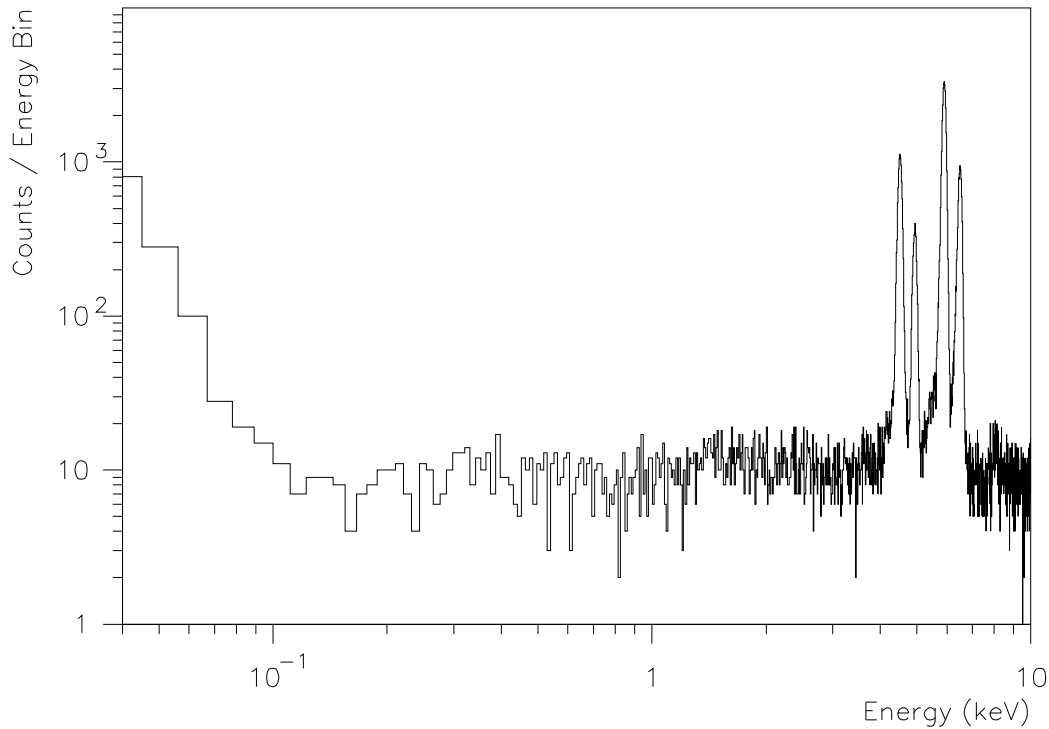,width=14cm}
\caption{
Measured spectrum with the ultra-low-energy
HPGe detector exposed to a $^{55}$Fe source.
A threshold of better than 100~eV is achieved.
}
\label{ulege}
\end{figure}

The case of ``Ultra Low-Energy'' ULB-HPGe detectors,
with the potential applications of 
Dark Matter searches  and 
neutrino-nuclei coherent scatterings, 
is being investigated. 
As depicted in the measurement with
a $^{55}$Fe X-ray source on a 5~g prototype
detector in Figure~\ref{ulege}, 
a hardware energy threshold of better than
100~eV has been achieved.
The lower energy peaks are those due
to back-scattering of the X-rays from Ti.
It is technically feasible to build an array of
such detectors to increase the target size to
the 1~kg  mass range.

\subsection{Dark Matter Searches with CsI(Tl)}

\begin{figure}
\center
{\bf a)}
\epsfig{file=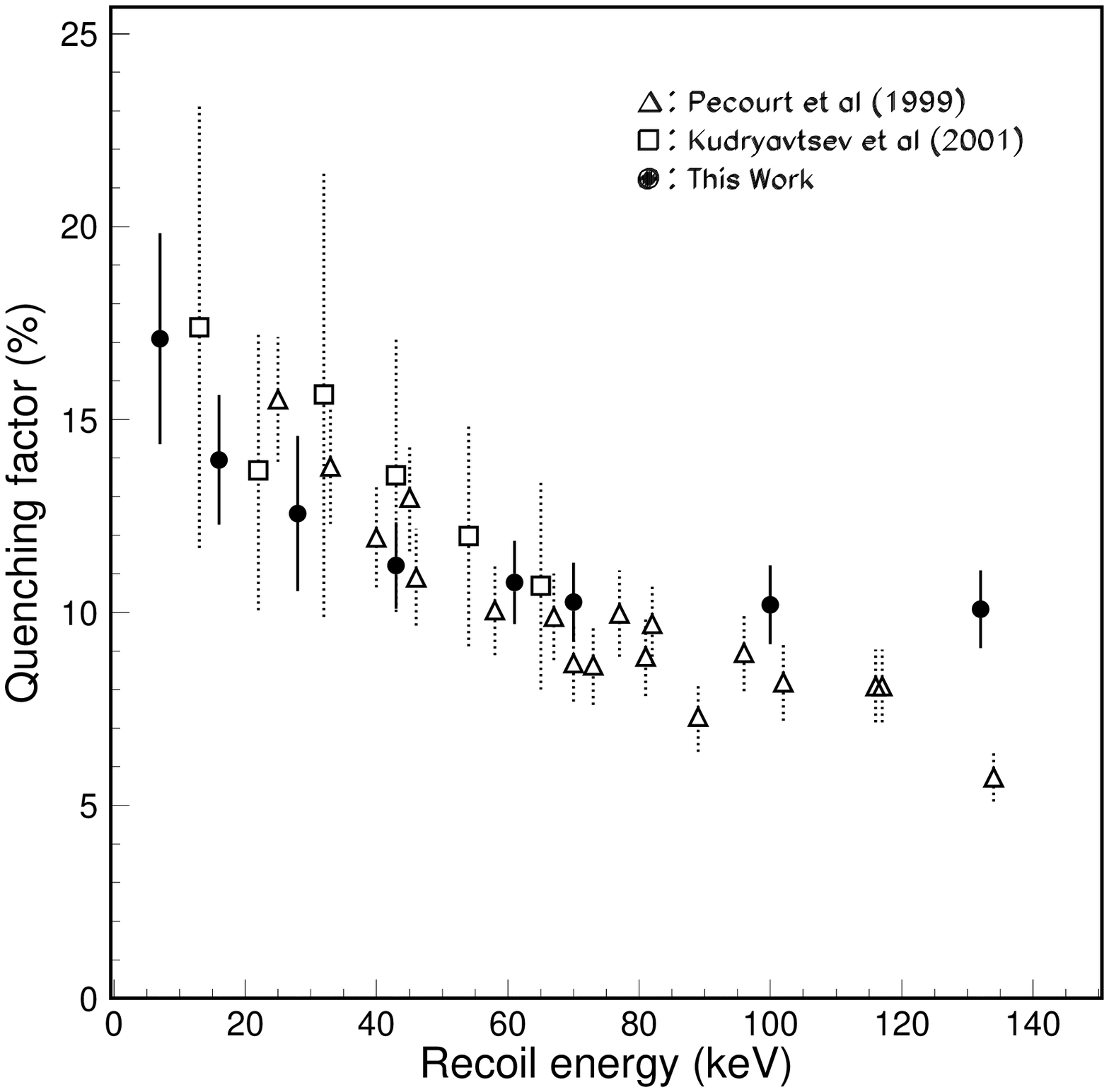,width=6.7cm}
{\bf b)}
\epsfig{file=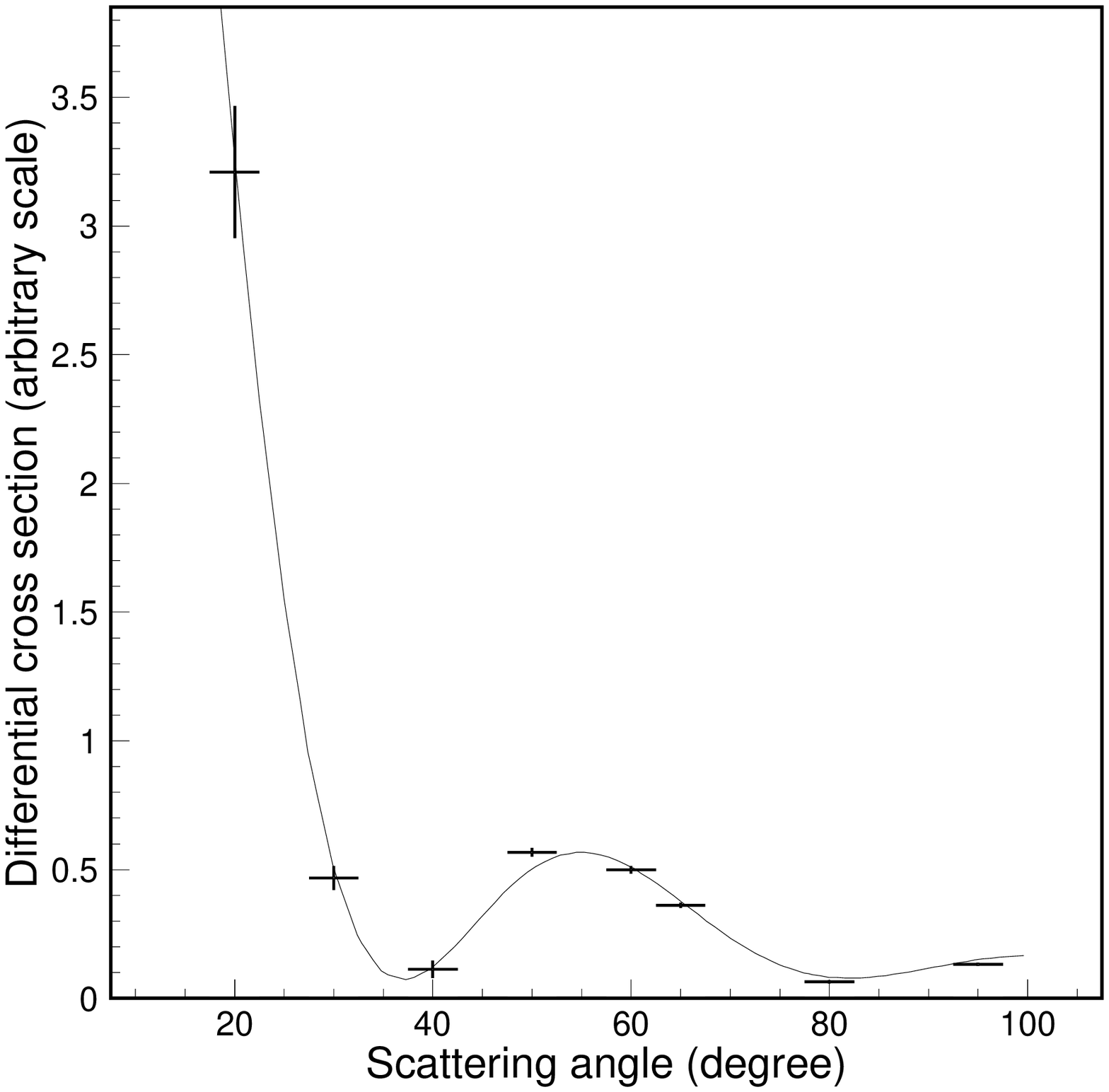,width=6.7cm}
\caption{
(a) The quenching factors, shown as black circles,
measured at IAE Tandem, as compared to 
previous work.
(b) The measured nuclear recoil
differential cross sections
in CsI(Tl), superimposed with the Optical Model
predictions.
}
\label{ciaebeam}
\end{figure}

Experiments based on the mass range of 100~kg of NaI(Tl)
are producing some of the most sensitive results in
Dark Matter ``WIMP'' searches\cite{naicdm}.
The feasibilities
and technical details of adapting CsI(Tl) or other
good candidate crystal like CaF$_2$(Eu) for WIMP Searches
have been studied.
A neutron test beam measurement for CsI(Tl)
was successfully performed at IAE 13~MV Tandem
accelerator\cite{nbeam}.
We have collected the lowest threshold data
for nuclear recoils in CsI(Tl), enabling us to
derive the quenching factors, displayed
in Figure~\ref{ciaebeam}a, as well as
to study the pulse shape discrimination
techniques at the realistically 
low light output regime\cite{lepsd}. 
The measurements also provide
the first confirmation of the Optical Model
predictions on neutron elastic scatterings with a
direct measurement on the nuclear recoils of heavy nuclei,
as illustrated by the differential cross-section
measurements of Figure~\ref{ciaebeam}b.
A full scale Dark Matter experiment
with CsI(Tl) crystals is being pursued
by the KIMS Collaboration 
in South Korea\cite{kims}.

\subsection{Radio-purity Measurements with Accelerator Mass Spectrometry}

Measuring the radio-purity of detector target materials
as well as other laboratory components are crucial to
the success of low-background experiments.
The typical methods are direct photon counting with
high-purity germanium detectors, $\alpha$-counting
with silicon detectors, conventional
mass spectrometry or the neutron
activation techniques.
We are exploring the capabilities
of radio-purity measurements further with the
new Accelerator Mass Spectroscopy (AMS) techniques\cite{amsradio}.
This approach may be complementary to existing
methods since it is in principle a superior
and more versatile method as demonstrated in
the $^{13}$C system, and it is
sensitive to radioactive isotopes that do
not emit $\gamma$-rays (like single beta-decays
from $^{87}$Rb and $^{129}$I) or where $\gamma$
emissions are suppressed (for instance,
measuring $^{39}$K provides a gain of
10$^5$ in sensitivity relative to detecting
$\gamma$'s from $^{40}$K).
A pilot measurement of the $^{129}$I/$^{127}$I ratio
($< 10^{-12}$) 
in CsI was successfully performed
demonstrating the capabilities
of the Collaboration. Further beam time
is scheduled at the IAE AMS facilities\cite{ciaeams}
to devise measuring schemes for the other
other candidate isotopes like
$^{238}$U, $^{232}$Th, $^{87}$Rb, $^{40}$K 
in liquid and crystal scintillators
beyond the present capabilities by the
other techniques.
The first isotope to study is on $^{40}$K,
where the goal sensitivity of
a $\rm{10^{-14} g/g}$ should be
achieve-able by the AMS techniques.

\subsection{Upgrade of FADC for LEPS Experiment}

\begin{figure}
\center
\epsfig{file=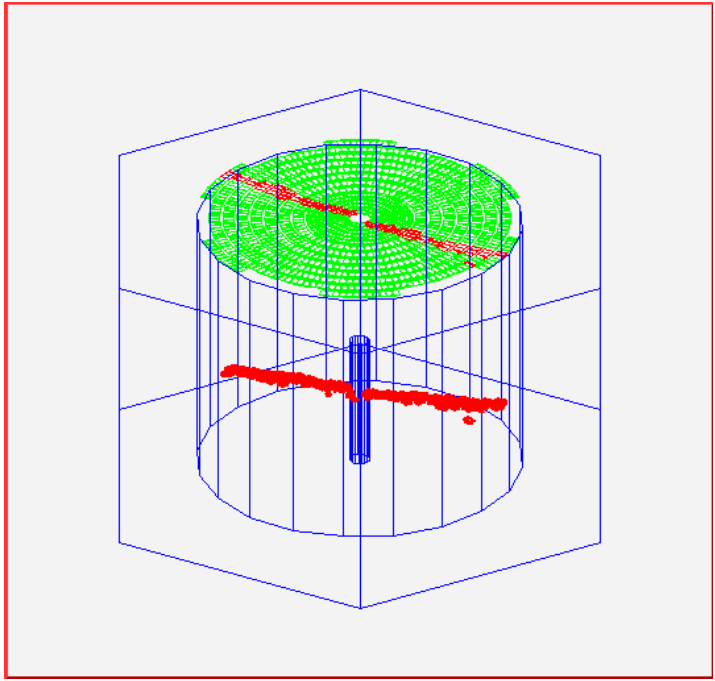,width=7cm}\\[3ex]
\caption{
A typical cosmic-ray event recorded by
the TPC for the LEPS Experiment,
using an FADC-based data acquisition system. 
}
\label{fadctpc}
\end{figure}

Based on the design and operation
of the FADCs at the KS experiment, 
we developed new FADCs
for the Time Projection Chamber (TPC)
constructed as a sub-detector for
the LEPS experiment at the SPring8 Synchrotron
Facilities in Japan\cite{Spring8}.
The LEPS FADCs
have 40~MHz sampling rate, 10-bit dynamic
range, 32 channels per module 
and are equipped with Field Programmable Gate Array (FPGA)
capabilities for real time data processing.
The TPC has 1000~readout channels
and a typical cosmic-ray event is depicted
in Figure~\ref{fadctpc}.
The new system will be commissioned at LEPS 
in summer 2003. The upgraded FADCs will
be further optimized and implemented 
to the KS reactor neutrino experiment
for data taking in 2004.

\section{Outlook}

The strong evidence of
neutrino masses and mixings\cite{pdg} lead to
intense world-wide efforts to pursue the next-generation
of neutrino projects. 
Neutrino physics and astrophysics will remain
a central subject in experimental particle physics
in the coming decade and beyond. There
are room for ground-breaking technical innovations $-$
as well as potentials for surprises in the scientific
results.

A collaboration among scientists from 
Taiwan and China has been built up 
with the goal of establishing
a qualified experimental program
in neutrino and astro-particle physics.
It is 
the first generation collaborative efforts
in large-scale basic research between scientists
from Taiwan and China.
The flagship effort is to
perform the first-ever particle
physics experiment in Taiwan
at the Kuo-Sheng Reactor Plant.
World-level sensitivities on the neutrino magnetic
moment and radiative lifetime have already been
achieved with the Period I data using a 
high-purity germanium detector.
Further measurements are pursued at the Kuo-Sheng
Laboratory, including
the Standard Model neutrino-electron 
scattering cross-section
as well as neutrino coherent scattering
with the nuclei.
A wide spectrum of R\&D projects are 
being pursued in parallel.

The importance of the implications and 
outcomes of the experiment and 
experience will
lie besides, if not beyond, neutrino physics.

\section{Acknowledgments}

The authors are grateful to the scientific members,
technical staff and industrial partners
of TEXONO Collaboration, as well as 
the concerned colleagues in our communities
for the many contributions which ``make 
it happen'' in such a short period of time.
Funding are
provided by the National Science Council, Taiwan
and 
the National Science Foundation, China, as well
as from the operational funds of the 
collaborating institutes.


\end{document}